\documentclass[conference]{IEEEtran}
\IEEEoverridecommandlockouts

\usepackage{cite}
\usepackage{amsmath,amssymb,amsfonts}
\usepackage{algorithmic}
\usepackage{graphicx}
\usepackage{textcomp}
\usepackage{xcolor}
\def\BibTeX{{\rm B\kern-.05em{\sc i\kern-.025em b}\kern-.08em
    T\kern-.1667em\lower.7ex\hbox{E}\kern-.125emX}}

\usepackage{array}
\usepackage[caption=false,font=normalsize,labelfont=sf,textfont=sf]{subfig}
\usepackage{stfloats}
\usepackage{url}
\usepackage{verbatim}
\usepackage{balance}
\usepackage{enumitem}
\usepackage{booktabs}
\usepackage{multirow}
\usepackage{makecell}
\usepackage{tabularx}
\usepackage{arydshln}
\newcolumntype{C}[1]{>{\centering\arraybackslash}m{#1}}

\begin{document}

\title{Solar-Blind Ultraviolet Communication: \\ Taxonomy and Experimental Characterization}

\author{\IEEEauthorblockN{Jian Shao}
\IEEEauthorblockA{\textit{School of Engineering} \\
\textit{The University of Edinburgh}\\
Edinburgh, UK \\
j.shao-7@sms.ed.ac.uk}
\and
\IEEEauthorblockN{Wasiu O. Popoola}
\IEEEauthorblockA{\textit{School of Engineering} \\
\textit{The University of Edinburgh}\\
Edinburgh, UK \\
W.Popoola@ed.ac.uk}
}

\maketitle

\begin{abstract}
Solar-blind ultraviolet communication (UVC) exploits near-zero solar background radiation and strong atmospheric scattering in the 200--280~nm band, enabling photon-level daytime communication and propagation modes beyond conventional line-of-sight (LOS) links. However, existing studies mainly focus on individual channel types, and a unified channel classification is still lacking. This paper presents a taxonomy of all four fundamental solar-blind UVC channels together with representative experimental validations and a comparative performance analysis. The proposed framework provides practical guidelines for selecting appropriate channel architectures for future solar-blind UVC systems.
\end{abstract}


\section{Introduction}\label{section_introduction}

\IEEEPARstart {O}{ptical} wireless communication (OWC) has emerged as a promising complementary technology to conventional radio frequency (RF) communication for alleviating spectrum congestion and electromagnetic interference (EMI) issues \cite{krishnamoorthy2025optical}. However, existing visible-light communication (VLC) and infrared communication (IRC) links generally rely on unobstructed line-of-sight (LOS) propagation and often require stringent beam pointing, acquisition, and tracking (PAT) \cite{hou2024tutorial}. Compared with the inherently wide-beam and obstacle-penetrating characteristics of RF systems, these requirements increase the system complexity of OWC and constrain its deployment flexibility.

Solar-blind ultraviolet communication (UVC) operating in the 200--280~nm band, however, offers several unique advantages that can partially overcome these limitations. First, due to strong atmospheric ozone absorption, solar radiation within the solar-blind band is virtually absent at ground level, enabling photon-level signal detection under outdoor daytime conditions \cite{shaw2006extending}. Second, the shorter wavelength of UV light results in substantially stronger Rayleigh scattering than visible or infrared wavelengths, facilitating non-line-of-sight (NLOS) links through atmospheric scattering \cite{siegel2004short}. Consequently, in addition to the typical collimated-beam narrow-field-of-view (FOV) LOS channel commonly employed in OWC systems, solar-blind UVC enables a wider range of propagation modes, including diffused-beam LOS, reflection- and scattering-mediated NLOS mechanisms.

Although most solar-blind UVC channel types have been previously investigated, existing studies typically focus on individual propagation mechanisms and system configurations \cite{maclure2022hundred, wang2025large, cao2021reflection, shaw2006extending}. Despite extensive research, the terminology and classification of solar-blind UVC channels remain inconsistent across the literature. As a result, a unified framework for channel identification and performance comparison remains largely absent. Furthermore, practical UVC links frequently involve multiple concurrent propagation mechanisms, such as LOS components coexisting with atmospheric scattering or reflection components accompanied by scattering contributions. The relative contributions of these channels vary significantly with link geometry, transmitter/receiver (Tx/Rx) configurations, and environmental conditions. Therefore, a systematic classification of basic UVC channel types is beneficial for interpreting experimental results and establishing consistent terminology between different studies.

In this paper, four fundamental solar-blind UVC channel types are first defined and categorized in Section~\ref{section_classification}. Then, Section~\ref{section_experiment} presents an experimental survey of these channels, covering demonstrations that approach or match the best performance reported in the literature. Section~\ref{section_performanceComparison} provides a systematic comparison and discussion of channel characteristics, including path loss, bandwidth, data-rate capability, alignment requirements, coverage, and blockage tolerance. Finally, conclusions are drawn in Section~\ref{section_conclusion}.

\section{Classification of Solar-Blind UVC Channels}\label{section_classification}

\begin{figure}[!htbp]
    \centerline{\includegraphics[width=0.48\textwidth, keepaspectratio=true]{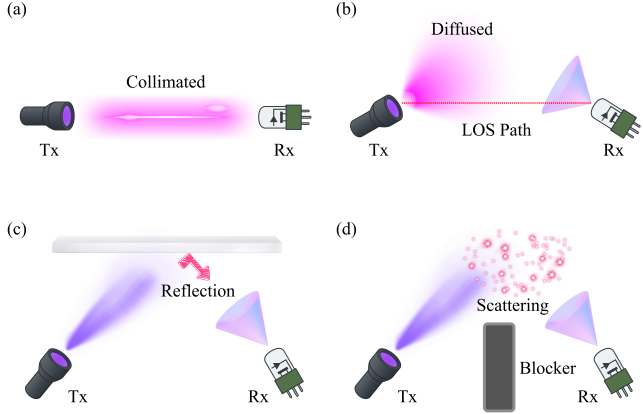}}
    \caption{Fundamental solar-blind UVC channel types: (a) collimated-beam narrow-FOV LOS channel, (b) diffused-beam wide-FOV LOS channel, (c) diffuse-reflection NLOS channel, (d) atmospheric-scattering NLOS channel.}
    \label{figure_ct}%
\end{figure}

Based on the Tx/Rx characteristics and the dominant photon propagation mechanisms, including LOS propagation, surface reflection, and atmospheric scattering, four basic solar-blind UVC channel types are defined in Fig.~\ref{figure_ct}, which represents the fundamental building blocks of practical UVC channels.

Figure~\ref{figure_ct}(a) illustrates the collimated-beam narrow-FOV LOS channel, which represents the most conventional OWC configuration employing optical beam collimation at Tx and optical concentration at Rx to maximize power efficiency and communication range \cite{maclure2022hundred}. In this configuration, laser or LED emissions are optically collimated into a narrow beam with a beam divergence angle (BDA) approaching $0^{\circ}$. At Rx, the received optical power is concentrated onto the detector sensing area through a convex lens, resulting in a narrow field of view (FOV) and high optical gain.

Figure~\ref{figure_ct}(b) shows the diffused-beam wide-FOV LOS channel. Although the link still relies on direct point-to-point LOS propagation, no optical collimation or concentration is employed at either Tx or Rx. The Tx typically exhibits a Lambertian radiation pattern of $I(\theta)=I_0\cos^n(\theta)$ ($0^\circ \le \theta \le 90^\circ$), where $I_0$ is the peak radiant intensity, $n$ is the Lambertian emission order, and $\theta$ is the emission angle measured from the Tx surface normal. Hence, the optical emission spreads throughout the whole hemisphere. Similarly, the Rx possesses a broad angular response and can detect incident photons with angle of arrival (AOA) ranging from $0^\circ$ to nearly $90^\circ$. Compared with the collimated-beam narrow-FOV LOS channel, this configuration offers substantially relaxed alignment requirements at the expense of reduced power efficiency.

In practice, many reported diffused-beam LOS UVC systems can be regarded as an intermediate configuration, referred to here as the diffused-beam narrow-FOV LOS channel \cite{vavoulas2019survey, yu2021experimental, wang2025large}. In such systems, the Tx emits a relatively wide beam without dedicated collimation optics, whereas the Rx employs optical concentrators, such as lenses or telescopes, to improve optical collection efficiency. Consequently, these systems exhibit characteristics that lie between the two fundamental types of LOS channel described above.

Generally, LOS UVC links provide relatively strong received optical signals compared with NLOS channels. Therefore, conventional optical detectors, such as photodetectors (PDs) and avalanche photodiodes (APDs), are often sufficient for direct optical intensity detection. For longer distances or highly attenuated links, photon-counting detectors, including photomultiplier tubes (PMTs) and single-photon avalanche diodes (SPADs), may be employed to improve Rx sensitivity.

Figure~\ref{figure_ct}(c) illustrates the diffuse-reflection NLOS channel. Unlike conventional OWC systems, which often rely on specular reflections from smooth surfaces, solar-blind UVC can also exploit weak diffuse reflections from a wide variety of rough natural and artificial surfaces, including indoor ceilings, tree canopies, and building facades, to establish photon-level communication links \cite{shao2025enhanced, cao2021reflection}.

Figure~\ref{figure_ct}(d) shows the atmospheric-scattering NLOS channel, in which emitted photons undergo single or multiple scattering events by atmospheric molecules and aerosols before reaching the Rx. It should be noted that many reported scattering-based NLOS UVC experiments employ a diffused-beam Tx together with a wide-FOV photon-sensitive Rx, while no explicit LOS blockage is introduced into the propagation path, which is essentially the same configuration as a diffused-beam wide-FOV LOS link \cite{xu2007experimental, du2022characteristic, pan2024beacon, deng2021realization}. Under such mixed-channel configurations, the received signal contains both direct LOS and scattered photons, as numerically proved in \cite{drost2013ultraviolet}. The relative contributions of the LOS and scattering NLOS components depend strongly on the link geometry, BDA/FOV, and pointing directions. For small to medium pointing misalignments, direct LOS propagation remains dominant, whereas the scattering component becomes increasingly significant as the Tx/Rx misalignment angle is further increased.

Therefore, when the objective is to investigate the characteristics of a purely scattering-mediated NLOS channel, the introduction of an explicit LOS blocker can be beneficial for eliminating residual LOS components and improving measurement accuracy. This approach allows the measured performance to be solely attributed to atmospheric scattering over the entire range of Tx/Rx orientations.

In general, both reflection- and scattering-mediated NLOS channels experience substantially higher path loss than LOS channels. Consequently, highly sensitive PMTs with large sensing areas are typically employed for signal reception.

\section{Representative Experimental Demonstrations}\label{section_experiment}

\subsection{Collimated- or Diffused-Beam Narrow-FOV LOS Channels}\label{subsection_cblos}

\begin{figure}[!hbp]
    \centerline{\includegraphics[width=0.475\textwidth, keepaspectratio=true]{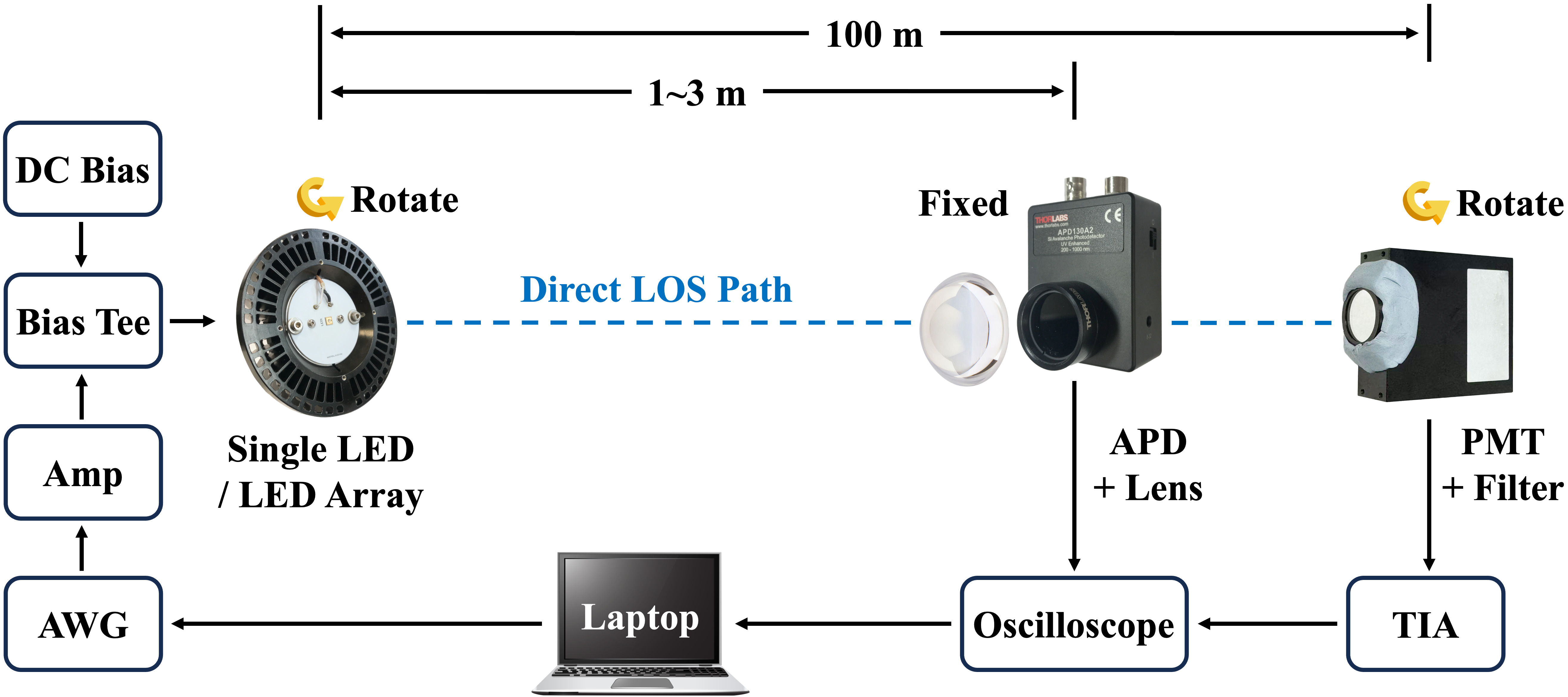}}
    \caption{System block diagram used for collimated-beam narrow-FOV LOS (APD), diffused-beam narrow-FOV LOS (APD), and diffused-beam wide-FOV LOS (PMT) links.}
    \label{figure_lbd}%
\end{figure}

The system block diagram demonstrating the collimated-beam narrow-FOV LOS UVC channel is included in Fig.~\ref{figure_lbd}. A single 265~nm, 100~mW LED (OSRAM SU CZHEF1.VC, 45~MHz bandwidth) served as the light source, while an APD (Thorlabs APD430A2/M) was used as the detector. Two convex lenses (Edmund Optics 84340) were placed on both sides to collimate and concentrate the beam. Intensity modulation/direct detection (IM/DD) with orthogonal frequency-division multiplexing (OFDM) was adopted.

\begin{figure}[!htbp]
    \centerline{\includegraphics[width=0.48\textwidth, keepaspectratio=true]{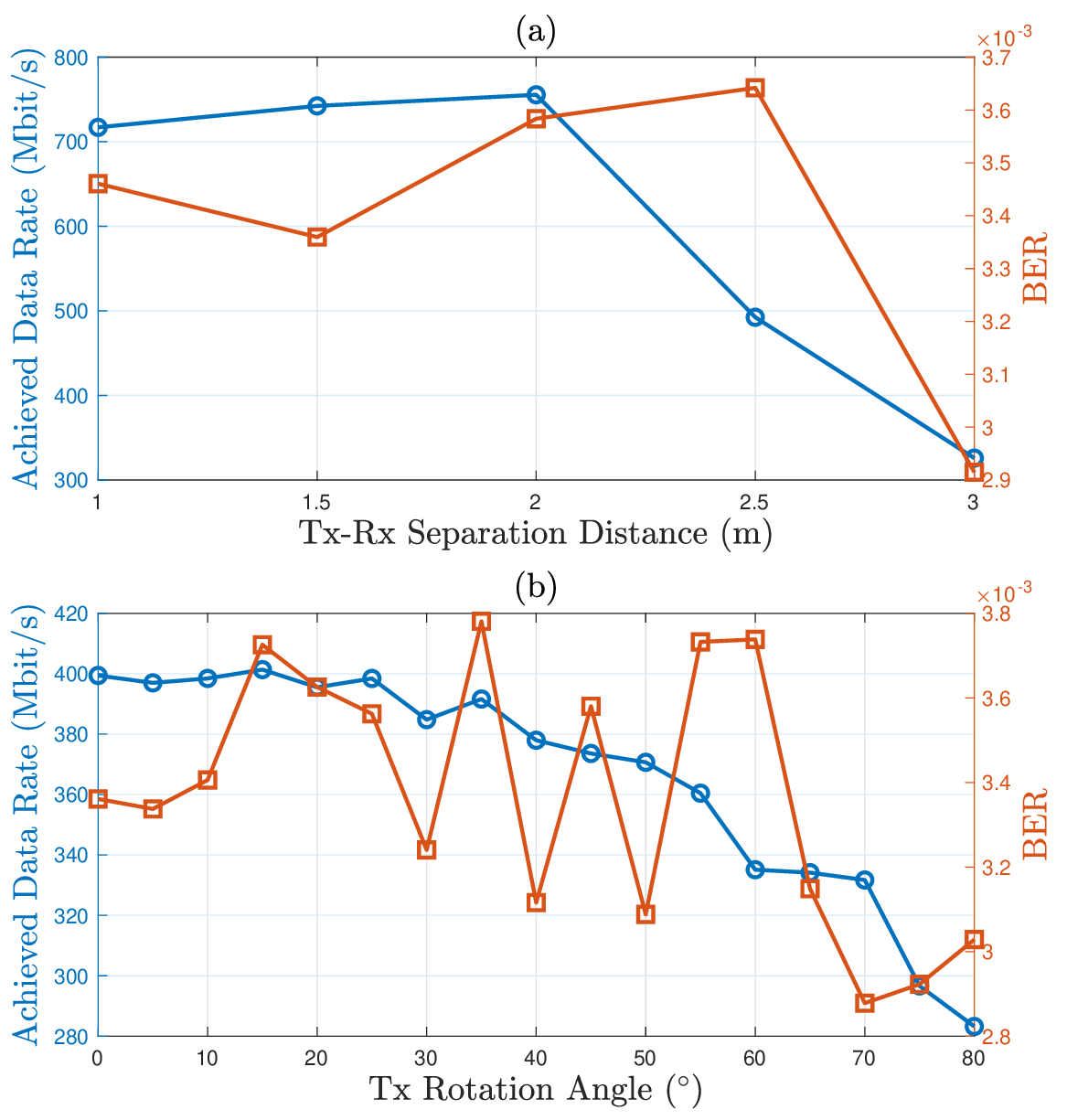}}
    \caption{(a) Achieved data rate and BER vs. Tx–Rx separation distance for a collimated-beam narrow-FOV LOS channel. (b) Achieved data rate and BER vs. Tx rotation angle for a diffused-beam narrow-FOV LOS channel.}
    \label{figure_nfl}%
\end{figure}

After optimizing the optical alignment and modulation parameters, the maximum achievable data rate was measured within the 7\% FEC threshold ($\mathrm{BER}=3.8\times10^{-3}$) at link distances of 1--3~m, as presented in Fig.~\ref{figure_nfl}(a). Data rates approaching 750~Mbps were achieved at link distances of 1--2~m. Beyond 2~m, the beam size expanded beyond the Rx lens aperture, introducing geometric loss and reducing the collected optical power. Consequently, the maximum achievable data rate gradually decreased to 326~Mbps at 3~m.

By removing the Tx lens while retaining the Rx lens, the system was converted into a diffused-beam narrow-FOV LOS UVC channel. The bare LED exhibited a full-width-at-half-maximum (FWHM) BDA of approximately $\pm60^\circ$. The link distance was fixed at 2.5~m, while the Tx rotation angle, denoted by $\theta_{\mathrm{Tx}}$, varied from $0^\circ$ to $80^\circ$. Owing to the rotational symmetry of the emission pattern, only one rotation direction was considered. The Rx remained aligned with the Tx because of its narrow FOV.

Figure~\ref{figure_nfl}(b) shows the maximum achievable data rate under the same 7\% FEC threshold. At $\theta_{\mathrm{Tx}}=0^\circ$, a data rate of 400~Mbps was achieved, lower than that of the collimated-beam configuration (approximately 500~Mbps) due to the reduced received optical power. As $\theta_{\mathrm{Tx}}$ increased, the received signal power gradually decreased, leading to a corresponding reduction in the achievable data rate. Nevertheless, even at $\theta_{\mathrm{Tx}}=80^\circ$, the link still supported a data rate of 283~Mbps, demonstrating the substantially improved angular coverage of the diffused-beam LOS channel.

\subsection{Diffused-Beam Wide-FOV LOS Channel}\label{subsection_dblos}

Figure~\ref{figure_lbd} also includes the system block diagram of the diffused-beam wide-FOV LOS channel. The link distance was increased to 100~m, and all optical collimation and concentration lenses were removed from both Tx and Rx. Due to severe path loss, the received optical power from single 0.1~W LED fell below the sensitivity limit of the APD. Therefore, a higher-power 0.9~W nine-LED array was employed as the Tx, while a photon-counting PMT (Hamamatsu H9305-09) was used as the Rx. This PMT exhibited a FWHM FOV of $\pm24^\circ$.

\begin{figure}[!htbp]
    \centerline{\includegraphics[width=0.485\textwidth, keepaspectratio=true]{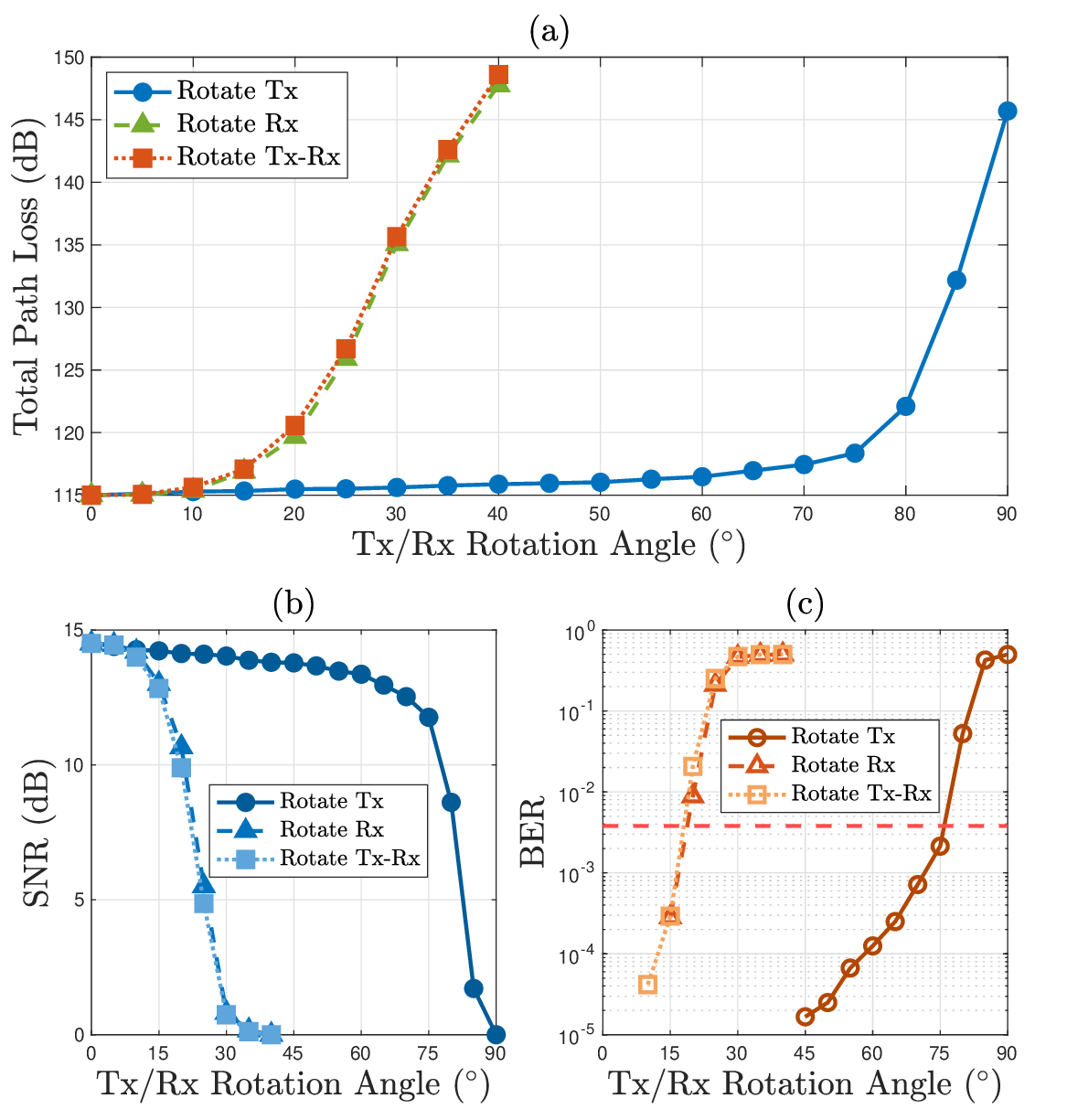}}
    \caption{Performance of a 100~m diffused-beam wide-FOV LOS UVC channel: (a) Total path loss vs. Tx/Rx rotation angle. (b) SNR vs. Tx/Rx rotation angle. (c) BER vs. Tx/Rx rotation angle. (*The red dashed line denotes the 7\% pre-FEC threshold of $\mathrm{BER}=3.8\times10^{-3}$.)}
    \label{figure_wfl}%
\end{figure}

Three representative misalignment scenarios were investigated: (a) Tx rotation with a fixed Rx, (b) Rx rotation with a fixed Tx, and (c) simultaneous Tx-Rx rotation away from each other. The measured total path loss is shown in Fig.~\ref{figure_wfl}(a). The results indicate that the path loss remains below 120~dB provided that the Tx and Rx rotation angles are within $77^\circ$ and $19^\circ$, respectively.

To evaluate communication performance, a 1.2~Mbps 8-level frequency-shift keying (8-FSK) signal was transmitted. The measured SNR and BER are presented in Fig.~\ref{figure_wfl}(b) and (c). Consistent with the path-loss results, reliable communication was maintained below the 7\% forward error correction (FEC) threshold ($\mathrm{BER}=3.8\times10^{-3}$) for Tx/Rx rotation angles up to approximately $76^\circ$ and $18^\circ$, respectively. These results demonstrate a significantly improved angular coverage of the diffused-beam wide-FOV LOS channel compared to conventional collimated-beam narrow-FOV configurations.

Furthermore, regardless of whether optical collimation or concentration optics are employed, the three different LOS channels presented in Subsections~\ref{subsection_cblos} and~\ref{subsection_dblos} all exhibited a measured system bandwidth of approximately 45~MHz, which was limited by the LED modulation bandwidth. In practice, LOS propagation itself introduces negligible temporal dispersion; therefore, the achievable system bandwidth is fundamentally determined by the Tx and Rx hardware rather than by the atmosphere channel.

\subsection{Diffuse-Reflection NLOS Channel}\label{subsection_rnlos}

\begin{figure}[!htbp]
    \centerline{\includegraphics[width=0.47\textwidth, keepaspectratio=true]{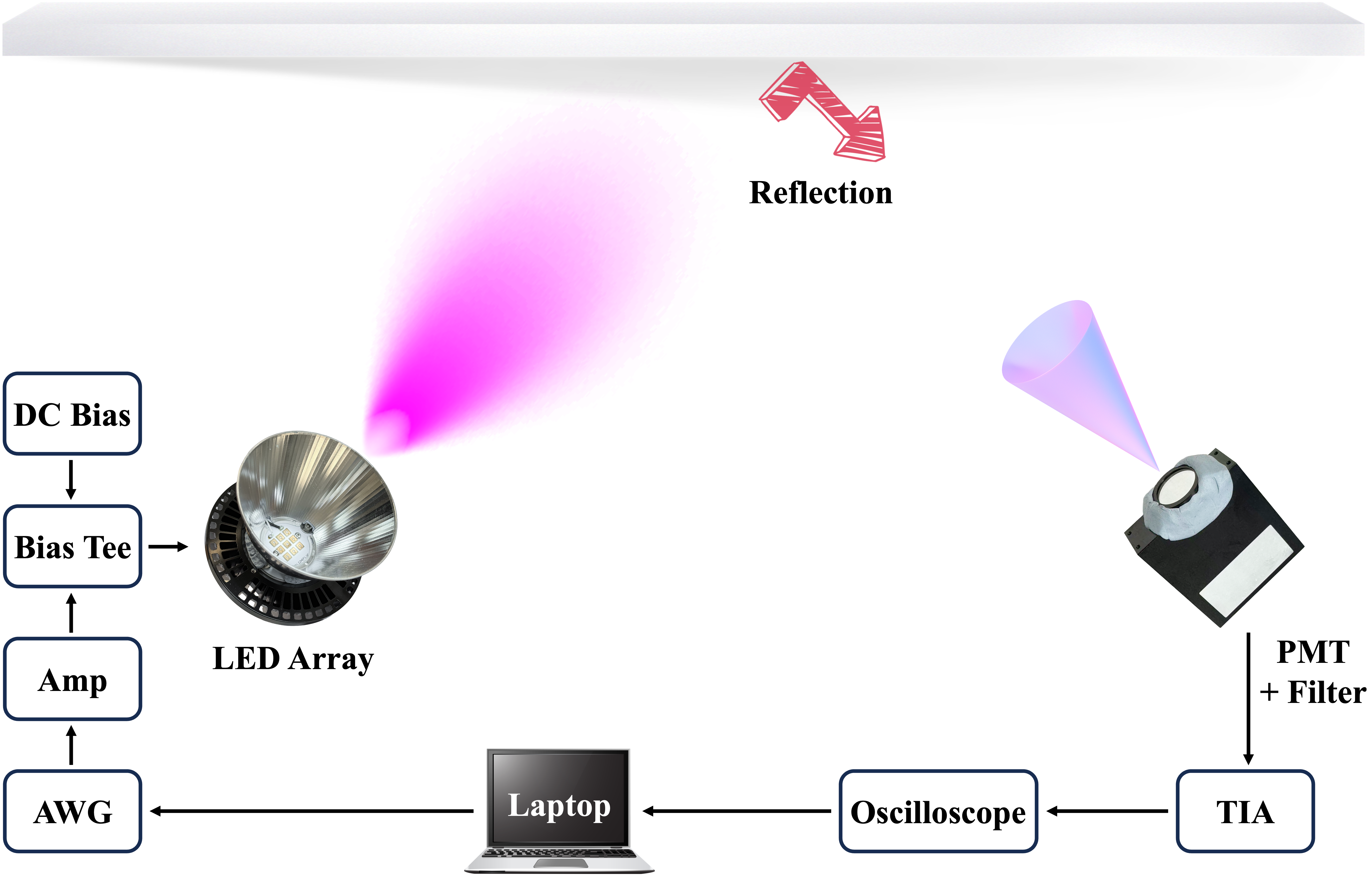}}
    \caption{System block diagram used for ceiling-reflection NLOS link.}
    \label{figure_rbd}%
\end{figure}

In our previous work \cite{shao2025enhanced}, a ceiling-reflection NLOS UVC system was experimentally demonstrated. The corresponding system block diagram is shown in Fig.~\ref{figure_rbd}. A nine-LED array was used to illuminate approximately 9~m$^2$ of the laboratory ceiling area. This ceiling acted as a secondary diffuse optical source, allowing the Rx to establish a reliable communication link by only roughly pointing toward the illuminated region.

Using on-off keying non-return-to-zero (OOK-NRZ) modulation, the system supported reliable data transmission at 2~Mbps. A notable advantage of the reflection-mediated NLOS channel is its strong resilience to blockage. Since communication relies on diffuse reflections rather than a single propagation path, blocking original reflecting surface does not necessarily interrupt the link. Instead, additional reflections are generated from the obstructing object itself, providing alternative propagation paths. Consequently, reliable communication can often be maintained unless Tx or Rx is completely covered. These results highlight the inherent spatial diversity and blockage tolerance of reflection-mediated NLOS UVC channels, making them attractive for wide coverage and obstacle-rich environments.

\subsection{Atmospheric-Scattering NLOS Channel}\label{subsection_snlos}

Atmospheric-scattering channels exhibit substantially higher path loss than other solar-blind UVC channel types. Therefore, particular care must be taken to eliminate contributions from stronger LOS or reflection components during measurements. For example, tree canopies, nearby buildings, or other surrounding objects can introduce certain reflected signals, while diffused-beam wide-FOV configurations without explicit LOS blockage retains significant LOS coupling.

\begin{figure}[!htbp]
    \centerline{\includegraphics[width=0.47\textwidth, keepaspectratio=true]{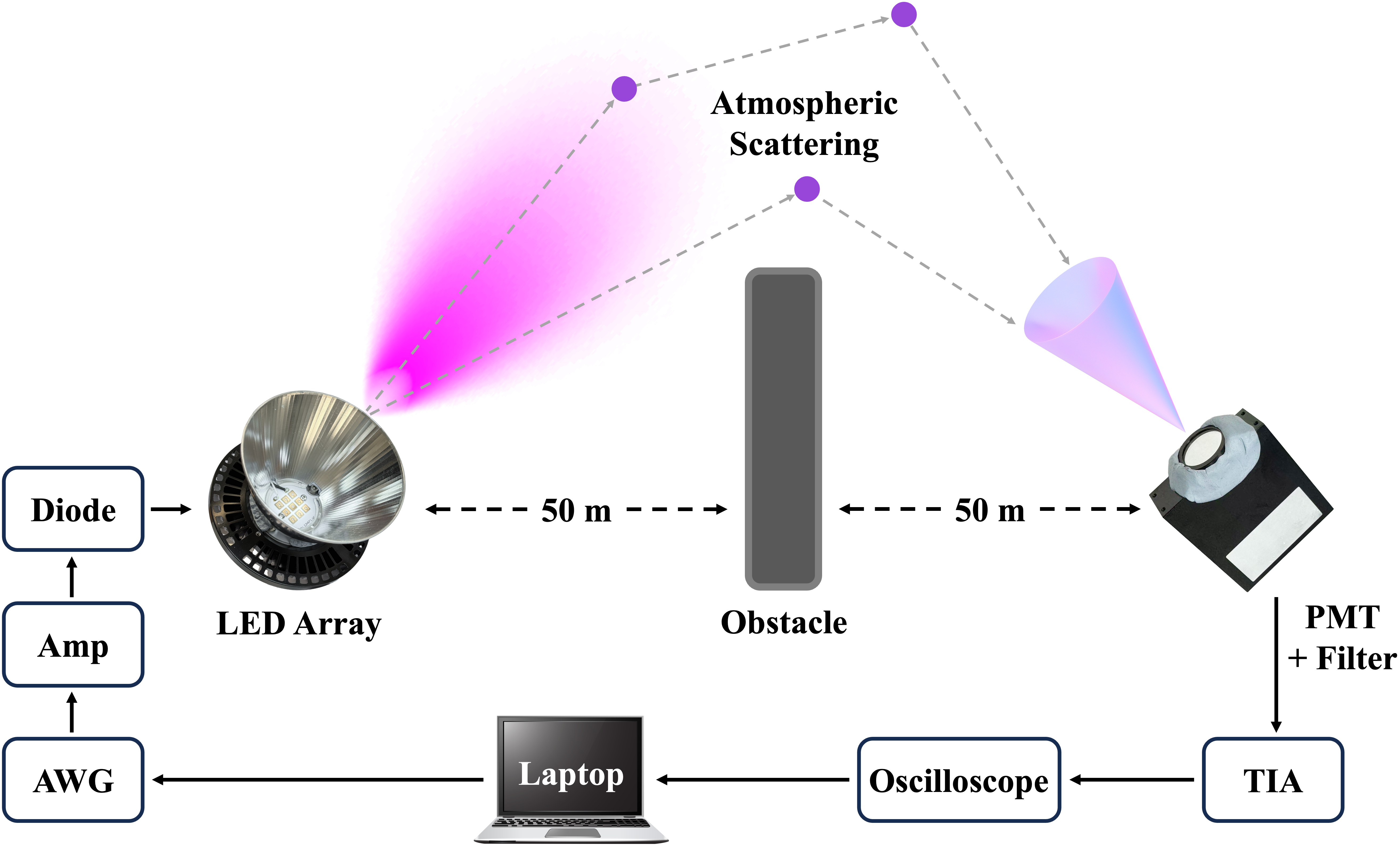}}
    \caption{System block diagram used for a pure scattering-mediated NLOS link.}
    \label{figure_sbd}%
\end{figure}

To investigate a purely scattering-mediated channel, experiments were conducted in an open parking area on the University of Edinburgh campus on the afternoon of May 8, 2026, under clear atmospheric conditions with few clouds and a meteorological visibility exceeding 10~km. As shown in Fig.~\ref{figure_sbd}, the Tx/Rx were separated by 100~m, with a 5-m-wide, 3-m-high opaque curtain positioned midway between them to completely block the LOS path. The Tx employed an 18-LED array with a peak optical power of approximately 1.8~W.

\begin{figure}[!htbp]
    \centerline{\includegraphics[width=0.485\textwidth, keepaspectratio=true]{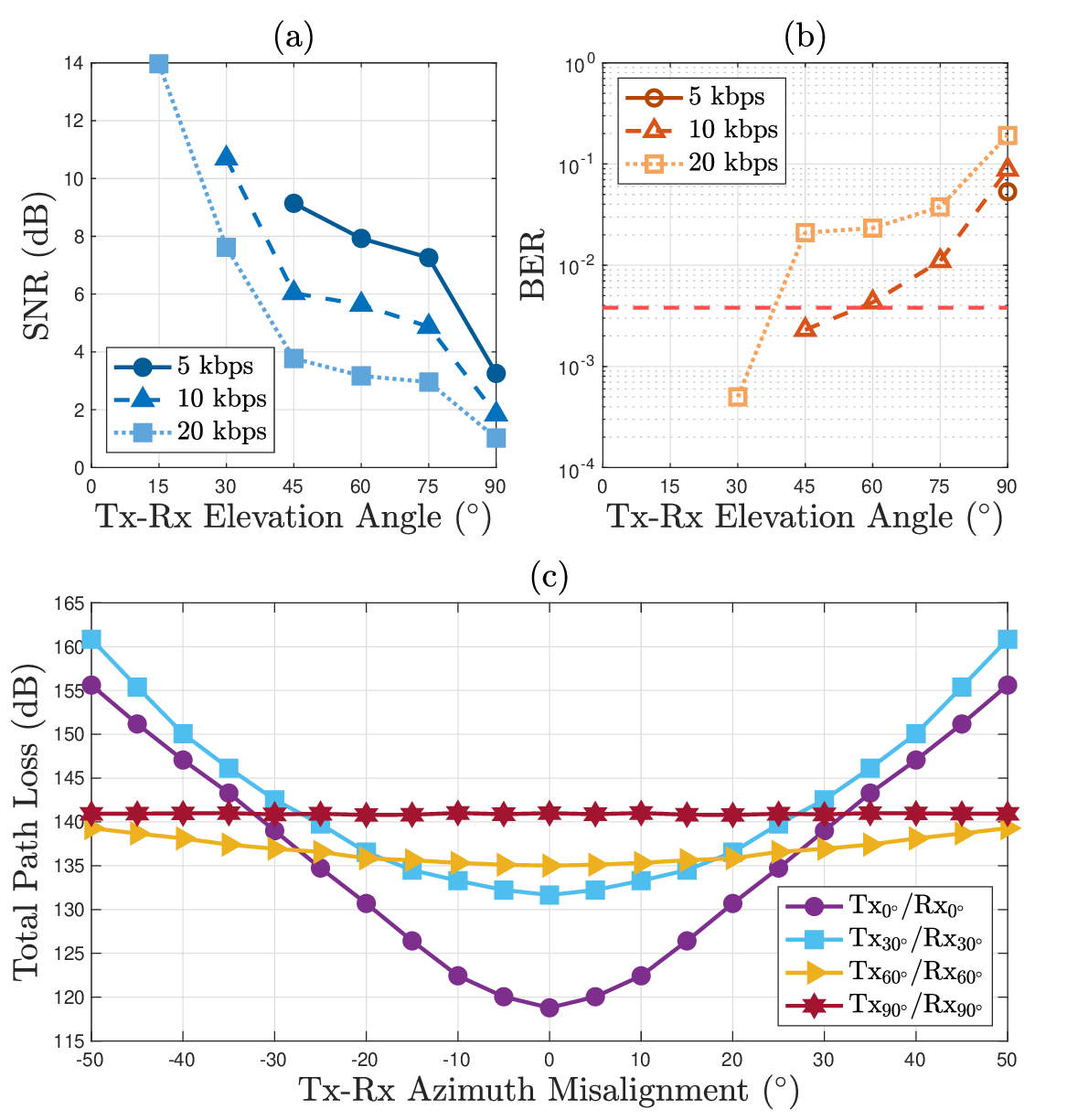}}
    \caption{Performance of a 100~m pure scattering NLOS channel with a 3~m obstacle: (a) SNR vs. Tx-Rx elevation angle. (b) BER vs. Tx-Rx elevation angle. (*The red dashed line denotes the 7\% pre-FEC threshold of $\mathrm{BER}=3.8\times10^{-3}$.) (c) Total path loss vs. Tx-Rx azimuth misalignment angle.}
    \label{figure_sn}%
\end{figure}

First, data transmission performance was evaluated under coplanar geometries. The Tx/Rx elevation angles were jointly increased from $0^\circ$ to $90^\circ$ in $15^\circ$ increments while maintaining zero azimuth misalignment. OOK-NRZ signals at 5, 10, and 20~kbps were transmitted. After photon-counting detection, the measured SNR and BER are presented in Fig.~\ref{figure_sn}(a) and (b). 

BER values below $10^{-4}$ are omitted as no bit errors were observed under the measurement resolution. When elevation angles increased, the scattering channel experienced higher path loss and smaller bandwidth. Consequently, the system supported reliable transmission at 20~kbps for low-elevation geometries, whereas only 5~kbps transmission remained feasible at high elevation angles. Despite the relatively modest data rate, the demonstrated performance is among the best reported for a strictly LOS-blocked scattering-mediated UVC link \cite{du2024implementation}.

The effect of azimuth misalignment was further investigated under non-coplanar geometries. Four Tx/Rx elevation angles ($0^\circ$, $30^\circ$, $60^\circ$, and $90^\circ$) were selected, and both Tx/Rx were simultaneously rotated in the same azimuthal direction from $0^\circ$ to $50^\circ$. The measured path-loss results are shown in Fig.~\ref{figure_sn}(c). For low-elevation geometries ($0^\circ$ and $30^\circ$), the initial path loss was relatively low but increased rapidly with azimuth misalignment. In contrast, high-elevation geometries ($60^\circ$ and $90^\circ$) exhibited higher baseline path loss but were considerably less sensitive to azimuth rotation.

Overall, the 100-m pure scattering-mediated UVC channel exhibited extremely high path loss exceeding 120~dB and supported significantly lower data rates than LOS or reflection-mediated channels. However, it offered the unique capability of maintaining reliable connection without any direct LOS path, highlighting its potential for highly obstructed communication scenarios.

\begin{table*}[b]
\centering
\caption{Comparison of Fundamental Solar-Blind UVC Channel Types}
\label{tab:channel_comparison}

\scriptsize
\setlength{\tabcolsep}{2.2pt}
\renewcommand{\arraystretch}{1.35}

\begin{tabular}{
C{1.55cm}
C{1.25cm}
C{1.55cm}
C{1.75cm}
C{1.40cm}
C{1.45cm}
C{1.65cm}
C{1.70cm}
C{1.70cm}
C{1.95cm}
C{0.60cm}
}

\toprule

\textbf{Channel Type}
&
\textbf{*Path Loss}
&
\textbf{\makecell{*Bandwidth}}
&
\textbf{\makecell{*Data Rate}}
&
\textbf{Range}
&
\textbf{Coverage}
&
\textbf{\makecell{Mobility\\(Alignment)}}
&
\textbf{\makecell{LOS-Blockage\\Tolerance}}
&
\textbf{\makecell{Atmospheric\\Robustness}}
&
\textbf{\makecell{Potential\\Applications}}
&
\textbf{\makecell{Ref.}}
\\

\midrule

Collimated LOS
&
\makecell{5--40 dB}
&
\makecell{not channel\\limited}
&
\makecell{Up To Tb/s}
&
\makecell{Long\\(1s--100s km)}
&
\makecell{Point-To-\\Point}
&
\makecell{Very Low\\(Strict PAT)}
&
None
&
\makecell{Very Low\\(Turbulence/Air\\Sensitive)}
&
\makecell{Satellite\\Underwater}
&
\cite{maclure2022hundred, alkhazragi20202}
\\

\cdashline{1-11}

Diffused LOS
&
\makecell{60--120 dB}
&
\makecell{not channel\\limited}
&
\makecell{Typically\\1s--100s Mb/s}
&
\makecell{Medium\\(0.1s--1s km)}
&
\makecell{Beam-Pattern\\Dependent}
&
\makecell{Good\\(Rough Alignment)}
&
None
&
\makecell{Low-Moderate\\(Air-Quality\\Sensitive)}
&
\makecell{UAV Swarm\\Remote Control}
&
This Work
\\

\cdashline{1-11}

Reflection NLOS
&
\makecell{100--140 dB}
&
\makecell{Moderate\\(MHz)}
&
\makecell{Typically\\1s--10s Mb/s}
&
\makecell{Short\\(10s--100s m)}
&
\makecell{Refl.-Surface\\Dependent}
&
\makecell{Good\\(Rough Alignment)}
&
\makecell{High\\(Obstacle-\\Bypassing)}
&
\makecell{Moderate\\(Air-Quality\\Sensitive)}
&
\makecell{Industry\\Indoor IoT}
&
\cite{shao2025enhanced, kurosawa2023solar}
\\

\cdashline{1-11}

Scattering NLOS
&
\makecell{120--160 dB}
&
\makecell{Low-Moderate\\(kHz--MHz)}
&
\makecell{Typically\\10s--100s kb/s}
&
\makecell{Short\\(10s--100s m)}
&
\makecell{Full-Area\\Coverage}
&
\makecell{Excellent\\(Rough Alignment)}
&
\makecell{Excellent\\(Obstacle-\\Bypassing)}
&
\makecell{Excellent\\(Averaging/\\Enhancing)}
&
\makecell{Tactical\\Vehicular}
&
\cite{du2024implementation}
\\

\bottomrule

\end{tabular}

\vspace{1mm}

\begin{flushleft}
\footnotesize
\textit{Note:} *Path loss, *bandwidth, and *data rate values are representative figures for 100-m solar-blind UVC links, estimated from literature reports and our experimental demonstrations using commercially available hardware. Actual performance depends on system configuration, environmental conditions, and implementation details; therefore, the values are provided for qualitative comparison only.

\end{flushleft}

\end{table*}

\section{Performance Comparison and Discussion}\label{section_performanceComparison}

Based on the experimental studies presented in Section~\ref{section_experiment} and representative results reported in the literature, the key characteristics of the four fundamental solar-blind UVC channel types are summarized in Table~\ref{tab:channel_comparison}. A comparative discussion is provided here. The collimated-beam narrow-FOV LOS channel provides the highest power efficiency and therefore exhibits the lowest path loss among all channel types. Since propagation is dominated by a direct LOS path with negligible temporal dispersion, the channel imposes virtually no bandwidth limitation, allowing the achievable data rate to be primarily determined by Tx and Rx hardware. This channel is well suited for long-range, point-to-point communications. However, it requires stringent PAT, offers limited mobility, and is highly sensitive to LOS blockage, atmospheric attenuation (caused by absorption and scattering), and turbulence.

In addition, the diffused-beam wide-FOV LOS channel sacrifices power efficiency in exchange for enhanced angular coverage and mobility. Although geometric loss leads to higher path loss than in collimated-beam systems, communication still relies on a direct LOS path and therefore experiences little bandwidth degradation. The relaxed alignment requirements enable broad-area coverage and facilitate one-to-many networking scenarios. Such channels are attractive for applications of remote control and UAV swarm communications.

Moreover, the reflection-mediated NLOS channel is suitable for environments where LOS links are frequently obstructed. Although the diffuse reflection process from rough surfaces introduces absorption loss and multipath propagation, the channel offers improved blockage tolerance and relaxed alignment requirements. As a result, reflection-mediated links can provide reliable connectivity in cluttered environments, making them attractive for indoor IoT, industrial monitoring, and EMI-sensitive scenarios where RF communication is undesirable.

Finally, the scattering-mediated NLOS channel enables communication when neither LOS nor reflection paths are available. In clear atmospheric conditions, the channel suffers from extremely high path loss and significant multipath broadening effects, resulting in limited communication range and data rate. Nevertheless, atmospheric scattering provides wide-area coverage, high mobility, and misalignment tolerance. Furthermore, distributed scattering paths can mitigate the impact of turbulence through path averaging \cite{yuan2024monte}. Under hazy or aerosol-rich conditions, enhanced scattering can substantially reduce path loss, thereby improving link availability under low-visibility conditions \cite{tang2023ultraviolet}. These unique properties make scattering UVC attractive for tactical vehicular communications and emergency communication deployment.

Overall, when a stable LOS path is available, the diffused-beam wide-FOV LOS channel is generally preferable for mobile nodes, as it provides relaxed alignment requirements while avoiding the extremely high path loss and limited bandwidth associated with scattering-mediated NLOS channels. If the LOS path is obstructed but suitable reflecting surfaces are available, a diffuse-reflection NLOS channel offers an effective alternative. Only when neither LOS nor reliable reflection paths are available should a scattering-mediated NLOS channel be considered. Moreover, the same hardware platform and modulation scheme can be readily shared among these channel types. This hierarchical channel selection strategy maximizes communication performance while minimizing system complexity. For example, in drone-to-drone and drone-to-ground communications, LOS connectivity is available in most practical scenarios. Consequently, LOS-based links are generally expected to provide superior overall communication performance, while scattering links are more suitable for vehicular applications where LOS and reflection paths cannot be continuously maintained.

\section{Conclusion}\label{section_conclusion}

This paper presents a unified taxonomy and experimental characterization of solar-blind UVC channels. Four fundamental channel types were systematically defined according to Tx/Rx characteristics and dominant photon propagation mechanisms. Representative experiments were demonstrated for each channel type, followed by a comparative evaluation of their system performances. The results show that each channel type offers distinct advantages and is suited to different application scenarios. Rather than relying on a single propagation mode, future solar-blind UVC systems should flexibly exploit LOS, reflection, and scattering according to environmental conditions and system requirements. For multi-user networks, these complementary channel types can be dynamically selected or jointly exploited to provide flexible connectivity for users with different link geometries and blockage conditions. The proposed taxonomy and comparative analysis provide a unified framework for understanding and designing future solar-blind UVC systems.

\bibliographystyle{IEEEtran}
\bibliography{References}

\end{document}